\documentclass[showkeys,nofootinbib]{revtex4-2}

\usepackage[utf8]{inputenc} 
\usepackage[T1]{fontenc} 
\usepackage[english]{babel} 
\usepackage{xcolor} 
\usepackage{amsmath,amsfonts,amssymb,amsthm,braket,dsfont,comment,leftindex,bbm} 
\usepackage{graphicx} 
\usepackage{subcaption,ragged2e} 
\usepackage{tikz} 
\usepackage{multirow,afterpage,tabularray} 
\usepackage{fancyhdr,orcidlink} 
\usepackage[capitalise]{cleveref} 
\usepackage{enumitem} 
\usepackage{titlesec} 
\usepackage{lipsum}
\usepackage{listings} 
\usepackage{cancel}

\newcommand{\diff}{\ensuremath{\mathrm{d}}}
\newcommand{\bal}{\begin{aligned}}
\newcommand{\eal}{\end{aligned}}

\definecolor{ao(english)}{rgb}{0.0, 0.5, 0.0}

\renewcommand{\bm}[1]{\boldsymbol{#1}}

\titleformat{\subsubsection}
   {\normalfont\fontsize{9pt}{11pt}\selectfont\bfseries}
   {\thesubsubsection.}
   {1em}
   {\centering}
\titleformat*{\subsubsection}{\normalsize\itshape}

\begin{document}

\title{Handling the Cornell potential within the Lagrange-mesh method in momentum space}

\author{\surname{Cyrille} Chevalier \orcidlink{0000-0002-4509-4309}}
\email[E-mail: ]{cyrille.chevalier@umons.ac.be}
\affiliation{Service de Physique Nucl\'{e}aire et Subnucl\'{e}aire,
Universit\'{e} de Mons,
UMONS Research Institute for Complex Systems,
Place du Parc 20, 7000 Mons, Belgium}

\author{\surname{Joachim} Viseur \orcidlink{0009-0006-3348-4402}}
\email[E-mail: ]{joachim.viseur@umons.ac.be}
\affiliation{Service de Physique Nucl\'{e}aire et Subnucl\'{e}aire,
Universit\'{e} de Mons,
UMONS Research Institute for Complex Systems,
Place du Parc 20, 7000 Mons, Belgium}

\date{\today}

\begin{abstract}
This work presents an alternative methodology for computing potentials matrix elements within the Lagrange-mesh method in momentum space. The proposed approach extends the range of treatable potentials to include previously inaccessible cases, such as the Coulomb and linear interactions. It enables, in particular, an efficient and accurate treatment of the Cornell potential, which plays an important role in potential models for hadronic physics. The method is validated across a variety of systems, with special attention given to the representation of both momentum and position probability densities. 
\end{abstract}
\keywords{Lagrange-mesh method, quantum two-body problem, momentum representation}

\maketitle


\section{Introduction}
\label{sec:intro}

Time-independent Schrödinger-like equations are still abundantly used in different fields of physics, including the phenomenology of hadrons \cite{godf85,caps86,fulc94,buis05,buis07,math08,meld08,chev25}. In three dimensions, such equations, analytically solvable only for a few combinations of kinetic and potential energy operators, must be solved by resorting to numerical approximate methods. For two-body problems, the Lagrange-mesh method (LMM) is a good candidate, capable of efficiently and accurately handling various potentials and kinematics \cite{baye15,sema01,lacr12}. The method is based on an expansion in trial states which are chosen so that they vanish at all points of a particular mesh but one. In this way, using appropriate quadratures to evaluate the integrals, the numerical computation of matrix elements is drastically simplified.

The LMM has abundantly been studied in configuration space and gives very good results with high accuracy for many two-body systems. Specific examples of such applications in hadronic physics can be found in References~\cite{buis05,math08}. Yet, for many systems, a momentum space description becomes more relevant. This is especially true when exotic kinematics or momentum-dependent interactions are used~\cite{szcz96,llan00,agui11,godf85}. Momentum space also fits better with descriptions based on quantum field theories, the latter being naturally written using momentum eigenstates~\cite{chev25}. These arguments motivate the development of a LMM directly written in momentum space. A first attend was made by G. Lacroix et al. in Reference~\cite{lacr11,lacr12}. For systems interacting via Gaussian and Yukawa potentials, their results proved consistent to those obtained in configuration space. However, it turns out that their methodology is not numerically appropriate to manage the Coulomb and the linear potentials. This proves problematic while studying the phenomenology of hadrons, as the latter potentials are widely used in this field, especially through Cornell-like interactions~\cite{fulc94,math08,godf85,caps86}. The present paper develops an alternative methodology to handle such interactions within the LMM in momentum space. It is inspired by the way exotic kinematics are handled in the original LMM based on configuration space \cite{sema01}. 

Section \ref{sec:LMM} is devoted to the theoretical description of the LMM. In case the method would be used in other fields than that which motivates this work, interactions and kinematics are kept general. To keep the article self-contained, some results from References~\cite{sema01} and~\cite{lacr12} are reviewed. An initiated reader is invited to focus on the essential contribution of this work to the LMM, which is contained in Section~\ref{ssec:pot_ME} where the alternative methodology to compute potential matrix elements is introduced. Accuracy tests on a Coulomb system are proposed in Section~\ref{ssec:analytic_test}. In Section~\ref{sec:conf_space}, the wave functions in position representation are obtained from those found in momentum space. Finally, Section~\ref{sec:app_to_mesons} illustrates that the alternative methodology indeed solves the initial problematic by reproducing the meson spectrum obtained in Reference~\cite{fulc94} while keeping a momentum space description. Let us also mention the \ref{app:LMM_direct_calc}, which illustrates why the methodology from Reference~\cite{lacr12}, is not appropriate to handle Coulomb and linear potentials. This work uses natural units ($\hbar=c=1$).


\section{The Lagrange Mesh Method}
\label{sec:LMM}

Let us consider a time-independent Schrödinger-like equation for a system of two bodies in their centre-of-mass frame,
\begin{equation}
    \left[T(p^2)+V(r^2)\right] \ket{\psi} = E \ket{\psi},
    \label{eq:eig_equation}
\end{equation}
Above, $p = |\bm{p}|$ and $r = |\bm{r}|$ stand for the modulus of the relative momentum and the separation of the two particles, respectively. The functions $T(p^2)$ and $V(r^2)$ represent a generic kinetic and potential contribution to the system's energy. The latter is assumed to depend only on the relative distance between the two particles\footnote{The present developments remain valid for any equation analogous to \eqref{eq:eig_equation} in which variables $r$ and $p$ have a different interpretation. The true requirement for $\bm{p}$ and $\bm{r}$ is to be conjugate variables.}.
To solve equation~\eqref{eq:eig_equation}, many resolution methods rely on variational computations. Namely, the eigenstates $\ket{\psi}$ are approximated by normalised linear combinations of a finite number of trial states $\ket{f_\alpha}$,
\begin{equation}
    |\psi\rangle \approx \sum_{\alpha=1}^N C_\alpha \ket{f_\alpha}\quad \text{with}\quad \sum_\alpha |C_\alpha|^2 = 1.
    \label{eq:expansion}
\end{equation}
For orthonormal trial states, MacDonald's theorem ensures that the above combinations provide good approximations of the true eigenvalues and corresponding eigenvectors  provided that the $C_\alpha$ parameters are solutions of the following eigenvalue equation \cite{macd33},
\begin{equation}
    \sum_{\beta=1}^N \bra{f_\alpha}\left[T(p^2)+V(r^2)\right]\ket{f_\beta} C_\beta = E\, C_\alpha.
    \label{eq:mat_eig_prob}
\end{equation}
If the above matrix elements are computed exactly, MacDonald's theorem even demonstrates that the approximate eigenenergies are upper bounds of the exact ones. The LMM suggests a definite set of trials states whose properties are tuned to facilitate the evaluation of $T(p^2)$ and $V(r^2)$ matrix elements by mean of Gauss-Laguerre quadratures.

Gaussian quadratures consist of approximating an integral by a weighted sum of its integrand evaluated in the zeros of definite orthogonal polynomials \cite{abra64}. Precisely, the $N$-points Gauss-Laguerre quadrature resorts to Laguerre polynomials of degree $N$, denoted $L_{N}(x)$. Using this quadrature, the integral of an arbitrary function of a single positive-definite variable $g(x)$ is approximated as
\begin{equation}
    \int_0^{\infty} g(x)dx \approx \sum_{k=1}^N \lambda_k g(x_k), 
    \label{eq:GL_quad}
\end{equation}
where $x_k$ are all the solutions of $L_N(x) = 0$. In the following, these roots are referred to as the Lagrange mesh. The weight coefficients $\lambda_k$ are chosen so that the above approximation is exact for $g(x)$ being any polynomial of order $2N-1$ multiplied by a decreasing exponential\footnote{It is worth mentioning that in a lot of references (see \cite{abra64}, for instance) the decreasing exponential is even made explicit at the level of equation \eqref{eq:GL_quad},
\begin{equation*}
    \int_0^{\infty} e^{-x} g(x)dx \approx \sum_{k=1}^N w_k g(x_k).
\end{equation*} One will easily convince that this results in a simple rescaling of the weight coefficients, $w_k = e^{-x_k} \lambda_k$. The present convention is chosen to simplify further expressions.}, $e^{-x}$. As a result, weights $\lambda_k$ are defined by the following formula \cite{lacr12},
\begin{equation}
    \ln \lambda_k = x_k - \ln x_k + 2 \ln \Gamma(N+1) - \sum_{j\neq i=1}^N 2\ln|x_k-x_j|.
\end{equation}

To define the trial set considered by the LMM, let us fix a representation for the states. In the current work, calculations are driven in momentum representations (the LMM in position representation has been abundantly developed in previous works \cite{baye15,sema01}). The following momentum wave functions are chosen for the trial states,
\begin{align}
     \braket{\boldsymbol{p}|f_i;lm} &= \frac{f_i(p/h)}{p\sqrt{h}} \tilde{Y}_{lm}(\theta,\phi)
     \label{eq:mom_repr}
\end{align}
with $i\in\{1,...,N\}$, $N$ being the number of mesh points, and with $\braket{\bm{p'}|\bm{p}} = \delta(p'-p)\delta(\cos\!\theta'-\cos\!\theta)\delta(\phi'-\phi)/p^2$ as the orthonormalisation convention for the momentum eigenstates. Above, $\theta$ and $\phi$ denote the polar and azimuth angles of the relative momentum $\bm{p}$, respectively. Angular momentum quantum numbers are provided to the trial states using modified spherical harmonic $\tilde{Y}_{lm}$ \cite{vars88}. In this work, a solution with a given angular momentum is always expected, but the procedure can easily be generalised to coupled channels. In definition~\eqref{eq:mom_repr}, the parameter $h$ is an energy scale whose job will be to adjust the mesh to the size of the system under consideration. The way this parameter is fixed is illustrated in an example in Section~\ref{ssec:analytic_test}. Finally, $f_j(x)$ are regularized Lagrange functions defined as follows \cite{lacr12},
\begin{equation}
f_i(x) = (-1)^{i}x_i^{-1/2} x(x-x_i)^{-1} L_N(x) e^{-x/2}.
\label{eq:lag_func}
\end{equation}
Note that this function vanishes at the origin, as expected for a physical wave function in the present conventions. In addition, each Lagrange function cancels at all points of the Lagrange mesh, except at a single one,
\begin{equation}
f_i(x_j) = \lambda_j^{-1/2} \delta_{ij},
\label{eq:lag_cond}
\end{equation}
This property is the keystone for constructing the LMM. As mentioned before, trial states~\eqref{eq:mom_repr} are expected to be orthonormalised. With the chosen orthonormalisation conventions for momentum eigenstates, their overlap writes as follows \cite{lacr12},
\begin{equation}
    \braket{f_i;lm|f_j;lm} = \int p^2\diff p \, \frac{f_i(p/h)}{p\sqrt{h}}\,\frac{f_j(p/h)}{p\sqrt{h}} \int \diff\!\cos\!\theta \diff\phi\,\tilde{Y}_{lm}^*(\theta,\phi)  \tilde{Y}_{lm}(\theta,\phi).
\end{equation}
The orthogonality of spherical harmonics \cite{vars88} simplifies the angular integral, leaving only the radial one,
\begin{equation}
    \braket{f_i;lm|f_j;lm} = \int_0^\infty \diff u\, f_i(u)f_j(u).
\end{equation}
where $u=p/h$. If one evaluates this integral with a $N$-points Gauss-Laguerre quadrature~\eqref{eq:GL_quad} and resorts to property~\eqref{eq:lag_cond}, the expected Kronecker delta is reproduced,
\begin{equation}
    \int_0^\infty \diff u\, f_i(u)f_j(u) \approx \sum_{k=1}^N \lambda_k f_i(x_k) f_j(x_k)= \delta_{ij}.
\end{equation}
The above calculation illustrates that the LMM trial states are orthogonal only at the Gauss approximation (represented here with an $\approx$ symbol). This observation is a first hint that the LMM is not a genuine variational method, as it evaluates its matrix elements approximately, resorting to Gauss-Laguerre quadratures. For instance, this makes the interpretation of its energies as upper bounds of the true eigenenergies impossible. Nevertheless, in spite of all these approximations, the method can achieve impressive accuracy levels~\cite{baye02,dohe17}.


\subsection{Matrix Element Computation in Momentum Space}
\label{sec:ME computation}

To apply equation~\eqref{eq:mat_eig_prob}, kinetic and potential energy matrix elements are now to be computed on trial states~\eqref{eq:mom_repr}. The next sections are devoted to these evaluations. It is shown that the strategy in reference~\cite{sema01} also holds in momentum space, provided that the treatment of kinetic and potential energy are inverted.

\subsubsection{Kinetic Energy Matrix Elements}

The kinetic energy matrix elements $\bra{f_i;lm}T(p^2)\ket{f_j;lm}$ are comparatively easier to determine. This operator being scalar, the angular part of the integral factorises and the normalisation of spherical harmonic can be used. Only the radial part of the integral remains,
\begin{equation}
\bra{f_i;lm}T(p^2)\ket{f_j;lm} = \int_0^{\infty} du \, T(h^2u^2) f_i(u) f_j(u)
\end{equation}
where $u = p/h$, again. As with the evaluation of the orthonormalisation of the trial states, a $N$-points Gauss-Laguerre quadrature~\eqref{eq:GL_quad} can be used to evaluated the residual integral. Resorting again to property~\eqref{eq:lag_cond}, one gets
\begin{equation}
    \bra{f_i;lm}T(p^2)\ket{f_j;lm} \approx T(h^2x_i^2) \delta_{ij},
\label{eq:kin_ME}
\end{equation}
With this formula, the evaluations of the kinetic energy $T$ at the mesh points are the only ones needed. This is a standard formula within the LMM in momentum representation \cite{lacr12}.

\subsubsection{Potential energy matrix elements}
\label{ssec:pot_ME}

Evaluating potential matrix elements prove more difficult. In Reference~\cite{lacr12}, a procedure based on the calculation of the potential's Fourier transform is suggested. Position-dependant potentials with Gaussian and Yukawa shapes are turned into their non-local momentum-dependant equivalents. Gauss-Laguerre quadratures are then used to obtain closed formulas to evaluate potential matrix elements. However, the strategy from \cite{lacr12} does not solve Coulomb or linear potentials, as illustrated in \ref{app:LMM_direct_calc}. For that reason, the current work suggests an alternative methodology for computing potential energy matrix elements which is compatible with the aforementioned potential shapes\footnote{Let us mention that similar calculations are presented in \cite{lacr12} to evaluate position-dependent observables. Nevertheless, such calculations have never been used to solve the Schrödinger-like equation~\eqref{eq:eig_equation} itself.}.

Basically, the potential energy matrix elements $ \bra{f_i;lm}V(r^2)\ket{f_j;lm}$ are calculated using the same four-step procedure as that used in reference \cite{sema01} to compute kinetic energy matrix elements in the configuration space. The resulting methodology is described below.
\begin{itemize}
\item First, the matrix elements of the operator $r^2$ are computed on the trial set, $R_{ij}^2 = \bra{f_i;lm}r^2\ket{f_j;lm}$. It is well-known that the expression of $r^2$ as a differential operator in momentum representation is the same as that of $p^2$ in position representation. Therefore, evaluating $r^2$ on momentum-dependant Lagrange functions necessarily uses the same formulas as evaluating $p^2$ on position-dependant Lagrange functions. The latter being computed in reference~\cite{sema01}, one simply has to mimic their calculation steps to show that
\begin{equation}
R_{ij}^2 \approx \frac{1}{h^2}\left(r_{ij}^2 + \frac{l(l+1)}{x_i}\delta_{ij}\right) \quad \text{with} \quad r_{ij}^2 =
\begin{cases}
(-1)^{i-j}(x_ix_j)^{-1/2}(x_i+x_j)(x_i-x_j)^{-2} &\text{for }i\neq j, \\
(12x_i^2)^{-1}\big( 4+(4N+2)x_i - x_i^2\big) &\text{for }i=j.
\end{cases}
\label{eq:R2}
\end{equation}
An approximate equality symbol is used because $N$-point Gauss-Laguerre quadratures are used to obtain the above expression.

\item The finite $R^2$ matrix is then diagonalised. The diagonal matrix is denoted $D^2$, while the corresponding transition matrix is denoted $S$,
\begin{equation}
R^2 = S D^{2} S^{-1}.
\label{eq:diag}
\end{equation}

\item The potential energy matrix $V_{D^{2}}$ is then computed on the basis of the $r^2$ eigenstates. This simply requires acting on each element of the diagonal matrix $D^2$ with the function $V$.
\item Finally, the potential matrix in the initial basis $V_{R^2}$ is obtained by returning to the original basis with the transformation law~\eqref{eq:diag},
\end{itemize}
\begin{equation}
    V_{R^2} = S V_{D^{2}} S^{-1}.
    \label{eq:VR2}
\end{equation}

As discussed in Reference~\cite{sema01}, the potential matrix elements obtained in this way are approximate for two reasons. First, formula~\eqref{eq:R2} using Gauss-Laguerre quadratures, the $R^2$ matrix elements are only evaluated approximately. But even if these matrix elements were computed exactly, as this methodology uses a finite number of trial states to picture $r^2$, the obtained eigenstates would only be approximations of the real ones. However, for both these approximations, increasing the number of mesh-points should result in a more accurate description. Note that, although the method was originally based on the variational theorems, the variational character of the solution is lost because all the matrix elements have been computed approximately.

\subsubsection{Resulting eigenvalues and eigenstates}
\label{sssec:acquisition}

Once the $N\times N$ potential and kinetic matrix elements are computed thanks to the above procedure and formula~\eqref{eq:kin_ME}, one can construct the matrix associated to $T(p^2)+V(r^2)$ and diagonalise it, as suggested by equation \eqref{eq:mat_eig_prob}. The resulting eigenvalues and eigenvectors provide approximations for the spectrum of the system. As only states with a given angular momentum $l$ have been used, the obtained approximations share this angular momentum. 

Concerning the eigenvectors, the $C_i$ coefficients are to be inserted into the equation~\eqref{eq:expansion} to infer the approximate eigenstate. In the next section, these will be used to compute approximate probability densities $\mathcal{P}\,\diff p$ associated with the modulus of the relative momentum $p$,
\begin{equation}
    \mathcal{P}(p)\diff p = p^2\diff p\int\diff\!\cos\!\theta\diff\phi\, \left|\braket{\bm{p}|\psi}\right|^2 \quad \text{where}\quad \ket{\psi} = \sum_{i=1}^N C_i \ket{f_i;lm}
\end{equation}
Using definition \eqref{eq:mom_repr}, one finds
\begin{equation}
    \left|\braket{\bm{p}|\psi}\right|^2 \approx \sum_{i,j=1}^N C_i^* C_j \frac{f_i(p/h)f_j(p/h)}{(p\sqrt{h})^2} \left|\tilde Y_{lm}(\theta,\phi)\right|^2.
\end{equation}
Since all matrix elements are real, the coefficients $C_i$ can likewise be chosen real. Using the orthonormality of the spherical harmonics \cite{vars88}, an expression for the approximate density is obtained,
\begin{align}
    \mathcal{P}(p)\diff p &\approx \frac{1}{h}\left(\sum_{i,j=1}^N C_iC_j f_i(p/h)f_j(p/h)\right) \diff p = \frac{1}{h}\left(\sum_{i=1}^N C_i f_i(p/h)\right)^2 \diff p.
    \label{eq:app_density}
\end{align}
The above formula can be used to represent the corresponding eigenstate or to compute momentum-dependent observables, for instance.


\subsection{Validation with analytical results}
\label{ssec:analytic_test}

\begin{figure}
    \centering
    \includegraphics[width=0.75\linewidth]{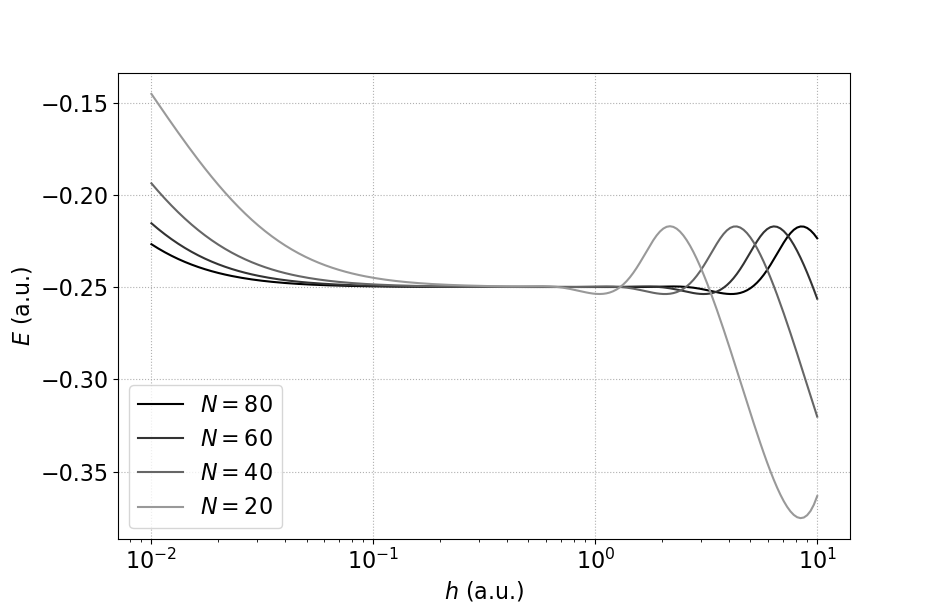}
    \caption{Ground-state energies of equation \eqref{eq:coul_ham} obtained with the LMM versus the scaling parameter $h$. Different mesh sizes are compared. The x-axis is plotted on a logarithmic scale.\justifying}
    \label{fig:h_dep}
\end{figure}

Now that the method is presented, it can be illustrated by comparison with analytical results. For this purpose, let us consider a system of two non-relativistic particles that interact with a pure Coulomb potential,
\begin{align}
   &T(p^2) = \frac{p^2}{2 \mu} \text{ with }\mu = \frac{m_1 m_2}{m_1+m_2}, &&V(r^2) =- \frac{a}{\sqrt{r^2}}.
\end{align}
For simplicity, particles are considered identical and have unit mass, namely $m_1=m_2=1$. The dimensionless parameter $a$ is also fixed at $1$. Arbitrary units (a.u.) are used. The Schrödinger equation to solve becomes 
\begin{equation}
    \left[p^2-\frac{1}{r}\right]\ket{\psi} = E\ket{\psi}.
    \label{eq:coul_ham}
\end{equation}
This equation can be solved exactly, its energy spectrum being below \cite{flug94,yane94},
\begin{equation}
    E_{n}=-\frac{1}{4n^2 } \quad \text{with }n = n_r + l +1.
    \label{eq:analytic_coulomb}
\end{equation}
Above, $n_r \in \mathbb{N}$ is the radial quantum number and $l \in \{0,...,n_r-1\}$. In the following, tests are performed on the three lowest states, denoted using spectroscopic notation $1S$, $2S$, and $1P$, respectively. The eigenstates of equation~\eqref{eq:coul_ham} can also be obtained analytically, even in momentum representation~\cite{flug94,yane94}. For the three lowest levels, the exact probability densities $\mathcal{P}\,\diff p$ associated with the modulus of the relative momentum $p$ are given by
\begin{subequations}\label{eq:anal_eigstates}
\begin{align}
&\mathcal{P}_{1S}(p)\, \diff p = \left(\frac{8}{\sqrt{\pi}} \frac{2p}{(1+4p^2)^2}\right)^2\diff p,\\
&\mathcal{P}_{2S}(p)\, \diff p = \left(2^5\sqrt{\frac{2}{\pi}} \frac{2p(1-16p^2)}{(1+16p^2)^3}\right)^2\diff p,\\
&\mathcal{P}_{1P}(p)\,\diff p = \left(2^7\sqrt{\frac{2}{3\pi}}\frac{4p^2}{(1+16p^2)^3}\right)^2\diff p.
\end{align}
\end{subequations}
These exact results can be compared to those obtained within the present LMM. First, the dependency in $h$ is investigated in Figure~\ref{fig:h_dep}. The ground-state energy is plotted versus $h$ for different mesh sizes. One can immediately see that the LMM is not variational as no clear minimum is observed. Instead, each curve shows a large plateau along which the energy is stable. The size of this plateau increases with $N$. This illustrates that the parameter $h$ enables to adapt the mesh to the typical scale of the system, but that an accurate value is not required as long as $N$ is chosen large enough. 

\begin{figure}
    \centering
    \includegraphics[width=0.5\linewidth]{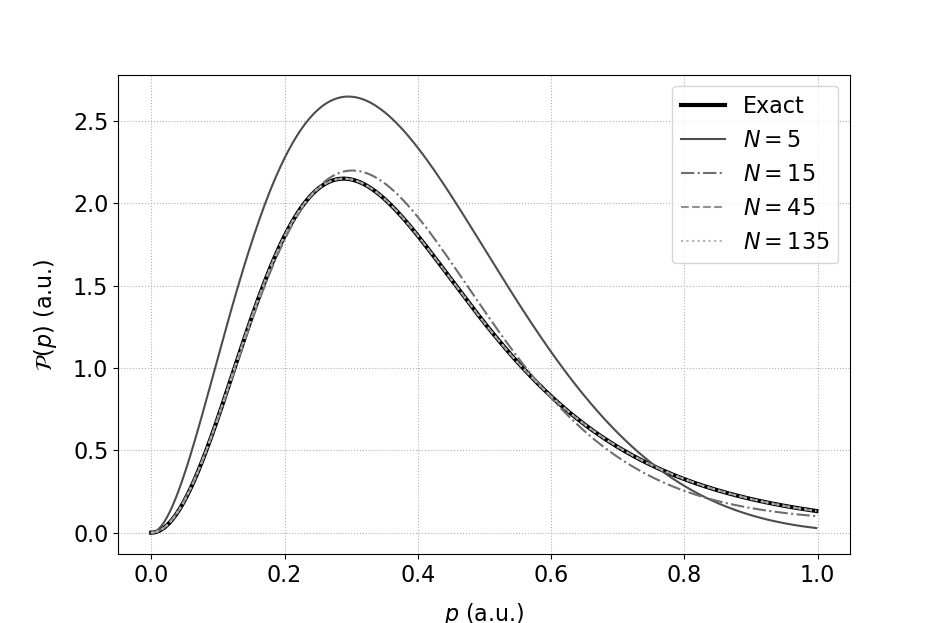}\includegraphics[width=0.5\linewidth]{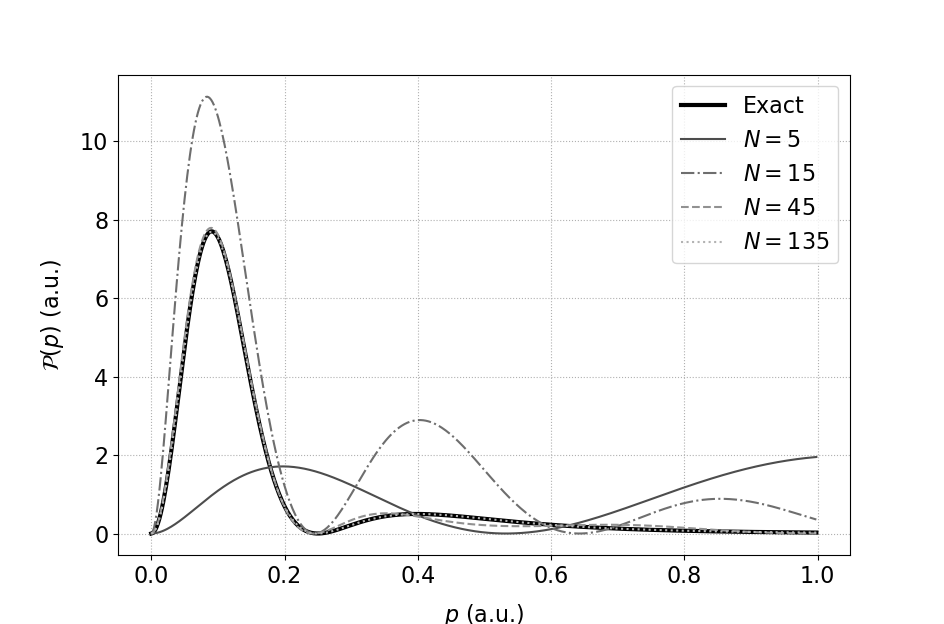}\\
    \includegraphics[width=0.5\linewidth]{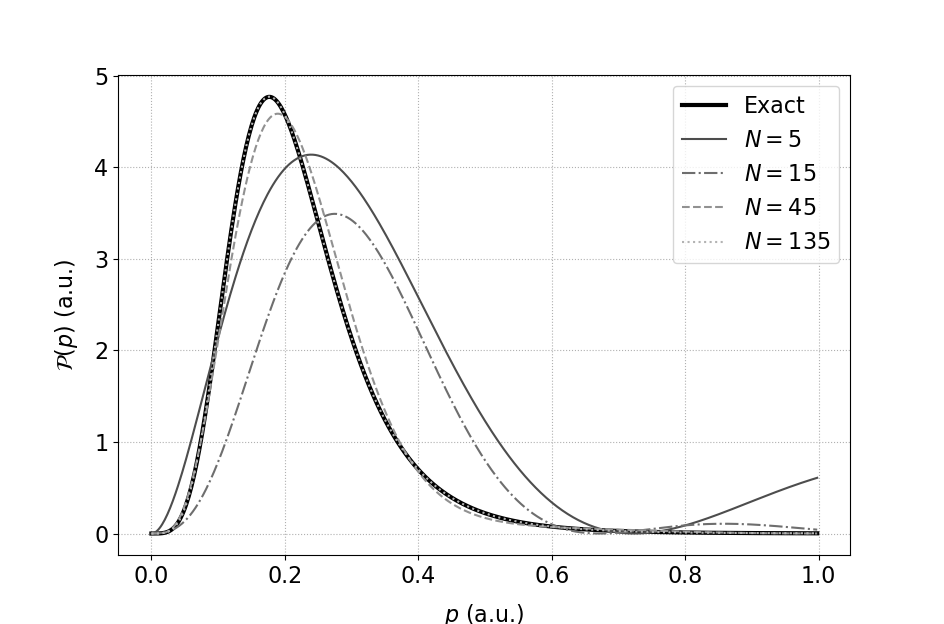}
    \caption{Representation of the probability densities $\mathcal{P}$ obtained with the LMM for the states $1S$, $2S$ and $1P$ (left, right and below, respectively) of equation \eqref{eq:coul_ham}. For each state, different mesh sizes are compared. Exact results are also displayed as thicker black lines. The LMM curves for $N=135$ are nearly indistinguishable from the exact ones. Calculations are performed with $h=0.5$. \justifying}
    \label{fig:coulomb_eig}
\end{figure}

Eigenenergies obtained with various mesh sizes $N$ and $h=0.1$ are displayed in Table~\ref{tab:val_coul} and compared with the exact ones. With only $50$ points, the LMM already reproduces at least three significant digits for all the investigated states. With $300$ points, the accuracy increases and more than five significant digits are systematically reproduced. Concerning eigenstates, Figure~\ref{fig:coulomb_eig} compares the probability density $\mathcal{P}(p)$ obtained using expression \eqref{eq:app_density} to the exact ones from~\eqref{eq:anal_eigstates}. Disparities decrease with increasing mesh size, with the two curves becoming indistinguishable for $N=135$. As an order of magnitude, our implementation on a regular laptop using Python 3 took only a few seconds to produces all the results from Table~\ref{tab:val_coul}.

\begin{table}
\begin{tabular}{lrrr}
\hline\hline
$N$& \hspace{2cm} $E_{1S}$ & \hspace{2cm} $E_{2S}$ & \hspace{2cm} $E_{1P}$\\
\hline
$50$ & $-0.249\,960\,128$ & $-0.062\,192\,468$ & $-0.062\,606\,365$ \\
$100$ & $-0.249\,989\,893$ & $-0.062\,495\,421$ & $-0.062\,501\,071$ \\
$200$ & $-0.249\,997\,451$ & $-0.062\,499\,681$ & $-0.062\,500\,000$ \\
$300$ & $-0.249\,998\,864$ & $-0.062\,499\,858$ & $-0.062\,500\,000$ \\
\hline
Analytical & $-0.25$ & $-0.0625$& $-0.0625$\\
\hline\hline
\end{tabular}
\caption{Test of the LMM for a system of two identical particles with unit mass and interacting via a Coulomb potential~\eqref{eq:coul_ham} for different mesh sizes $N$. Energies of the three lowest eigenstates are given in arbitrary units. The exact eigenenergies are given in the last line \cite{flug94,yane94}. Calculations are performed with $h=0.5$. \justifying}
\label{tab:val_coul}
\end{table}

Before moving on, let us mention that another test of the method has been conducted for a system of two relativistic particles interacting with a Gaussian potential. The corresponding results are summarised in \ref{app:gauss}. Since this system has also been solved in Reference~\cite{lacr12}, this test illustrates that the new methodology does not result in a poorer accuracy. 


\section{Representation of the state in configuration space}
\label{sec:conf_space}

The previous section closed with an illustration that the LMM in momentum space provides precise approximations for the momentum probability densities. However, the position probability density is required for some applications, notably within the field of hadronic physics \cite{buis07,meld08,tour25}. This suggests developing formulas to compute the Fourier transform of the LMM approximation,
\begin{equation}
    \braket{\bm{r}|f_i;lm} = \frac{1}{(2 \pi)^{3/2}} \int \diff^3p\, \frac{f_i(p/h)}{p\sqrt{h}} \tilde Y_{lm}(\theta,\phi)\, e^{i \bm{p}\cdot\bm{r}} 
\end{equation}
where the convention $\braket{\bm r|\bm p} = e^{i\bm{p}\cdot\bm{r}}/(2\pi)^{3/2}$ is used. Solving the angular dependence of the above integral results in
\begin{equation}
    \braket{\bm{r}|f_i;lm} = i^l \sqrt{\frac{2}{\pi}}\left(\int p^2\diff p\, \frac{f_i(p/h)}{p\sqrt{h}} \,j_l(pr)\right) \tilde Y_{lm}(\theta_r,\phi_r)   
\end{equation}
where $r$, $\theta_r$ and $\phi_r$ refer to as the modulus, orbital angle and azimuthal angle of the relative position between the particles, respectively. The $j_l$ function is a spherical Bessel function~\cite{abra64}. The remaining integral can be evaluated approximately by using again a $N$-point Gauss-Laguerre quadrature,
\begin{equation}
    \int p^2\diff p\, \frac{f_i(p/h)}{p\sqrt{h}} \,j_l(pr) \approx \sqrt{h^3\lambda_i}\, x_i\, j_l(hx_ir).
\end{equation}
As a result, one finds\footnote{Similar developments were carried out in Reference~\cite{lacr11} to transition from the position to the momentum representation. The inverse operation is addressed in the current work.} 
\begin{equation}
    \braket{\bm{r}|f_i;lm} = i^l \sqrt{\frac{2h^3\lambda_i}{\pi}} \, x_i\, j_l(hx_ir) \tilde Y_{lm}(\theta_r,\phi_r).
\end{equation}
Considering a given linear combination of trial states $\ket{f_i;lm}$ and following developments similar to those in Section~\ref{sssec:acquisition}, the radial probability density of the system, $\mathcal{R}(r)\diff r$, is given by
\begin{equation}
    \mathcal{R}(r)\diff r = \frac{2h^3}{\pi}\left(\sum_{i=1}^N C_i \sqrt{\lambda_i} \, x_ir\, j_l(hx_ir)\right)^2 \diff r.
    \label{eq:app_density_r}
\end{equation}

\subsection{Validation with analytical results}

\begin{figure}
    \centering
    \includegraphics[width=0.45\linewidth]{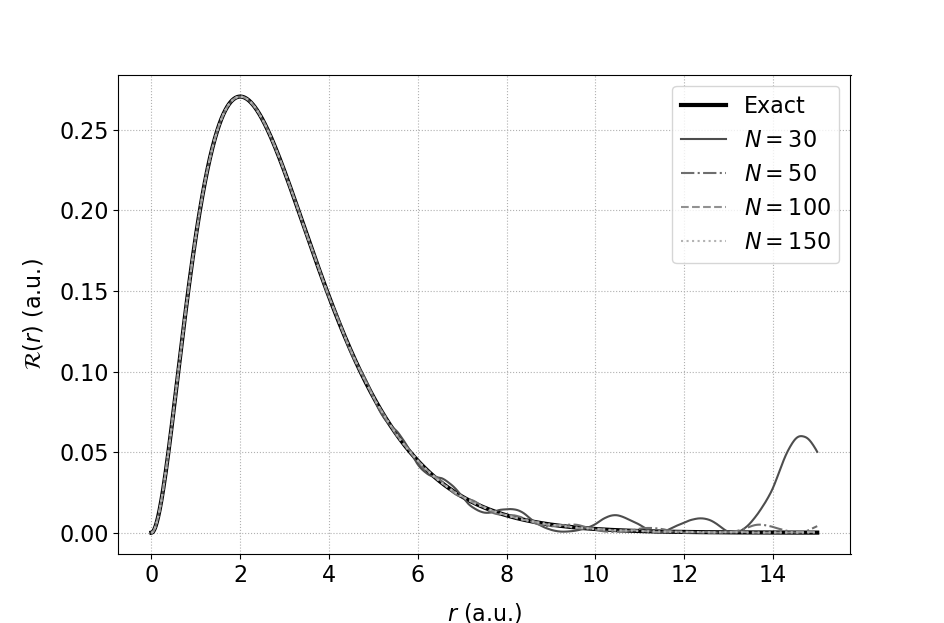} \includegraphics[width=0.45\linewidth]{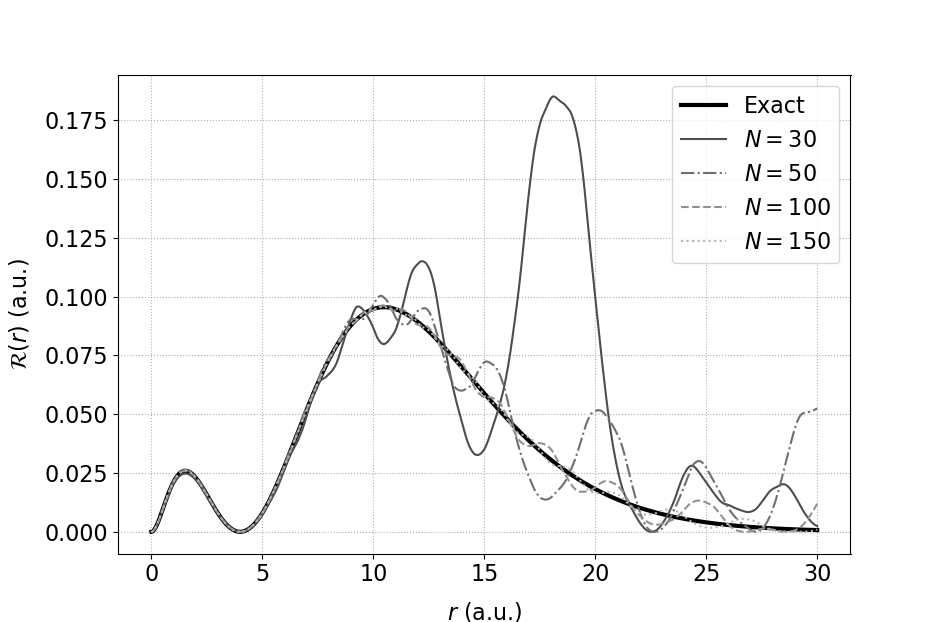} \\
    \includegraphics[width=0.45\linewidth]{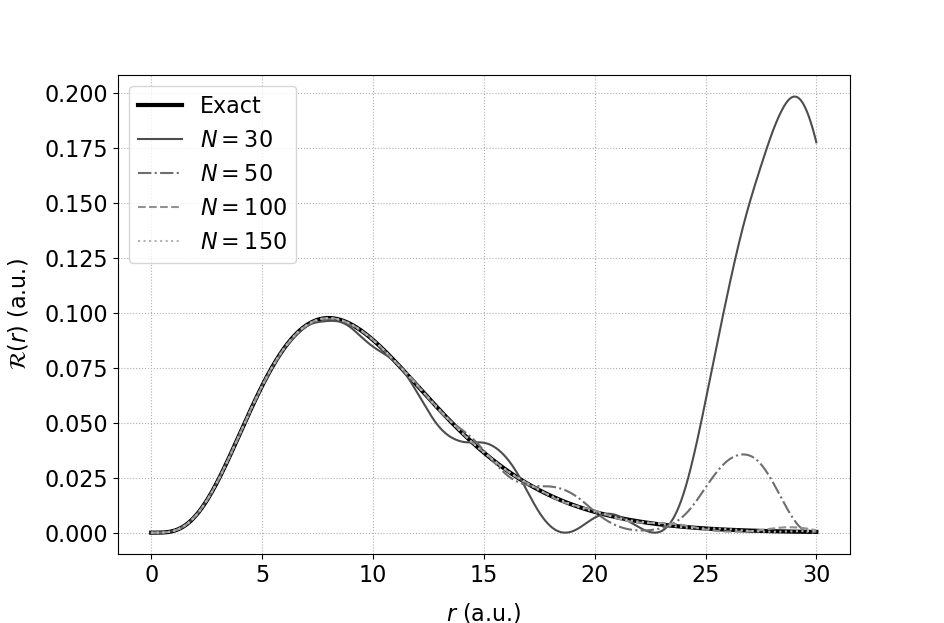}
    \caption{Representation of the probability densities $\mathcal{R}$ obtained with the LMM for the states $1S$, $2S$ and $1P$ (left, right and below, respectively) of equation \eqref{eq:coul_ham}. For each state, different mesh sizes are compared. Exact results are also displayed as thicker black lines. The LMM curves for $N=150$ are nearly indistinguishable from the exact ones. Calculations are performed with $h=0.5$. Note that the range of the abscissa differs for the $1S$ curve. \justifying}
    \label{fig:comp_densities_r}
\end{figure}

\begin{figure}
    \centering
    \includegraphics[width=0.45\linewidth]{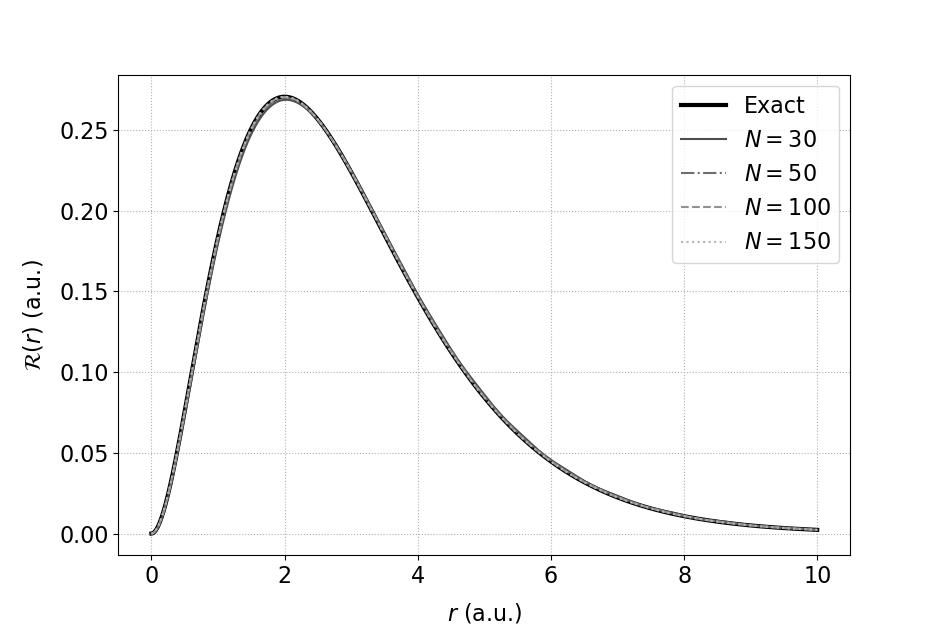} \includegraphics[width=0.45\linewidth]{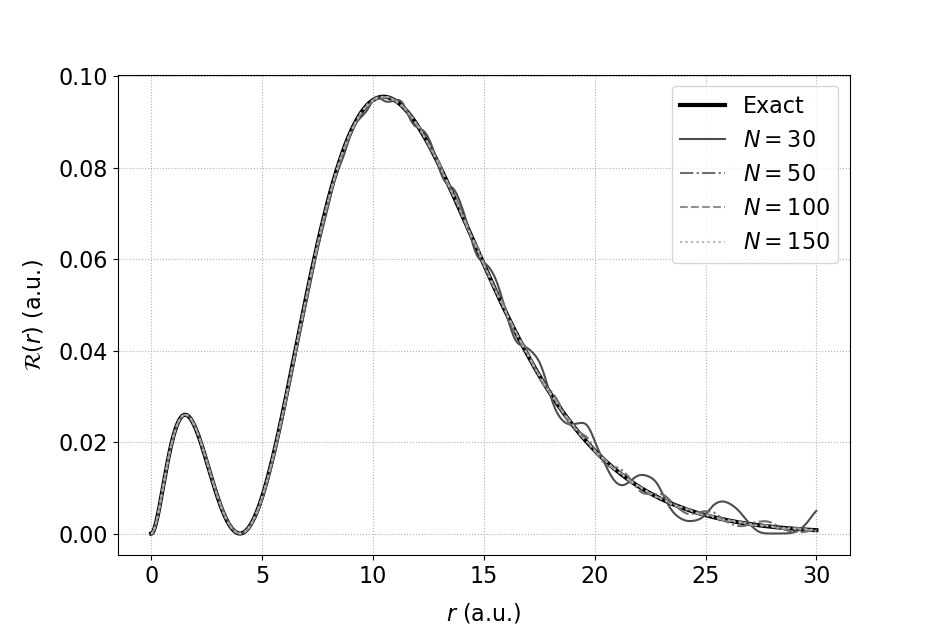} \\
    \includegraphics[width=0.45\linewidth]{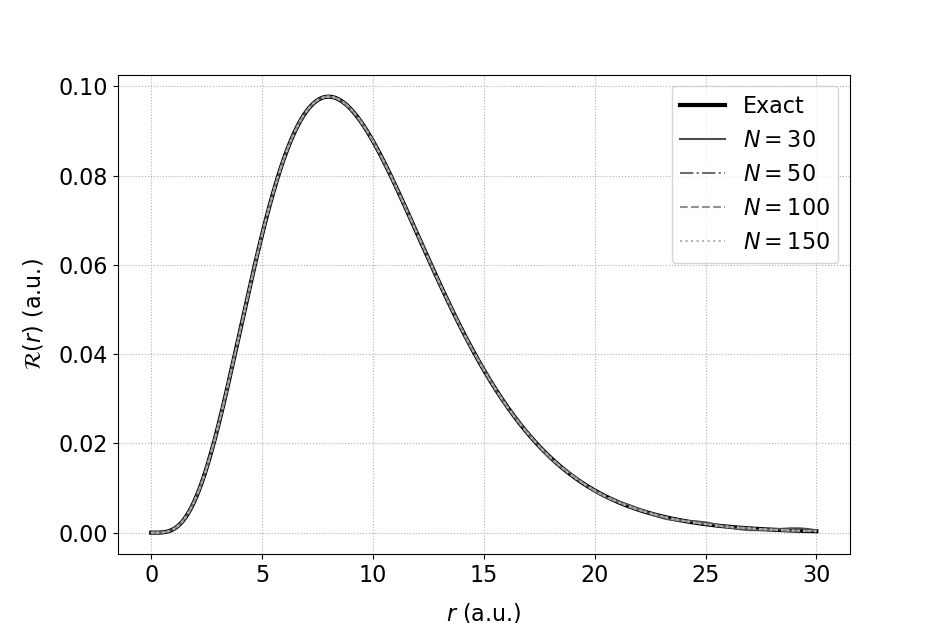}
    \caption{Representation of the probability densities $\mathcal{R}$ obtained with the LMM for the states $1S$, $2S$ and $1P$ (left, right and below, respectively) of equation \eqref{eq:coul_ham}. For each state, different mesh sizes are compared. Exact results are also displayed as thicker black lines. The LMM curves over $N=50$ are nearly indistinguishable from the exact ones. Calculations are performed with $h=0.1$. Note that the range of the abscissa differs for the $1S$ curve. \justifying}
    \label{fig:comp_densities_r_bis}
\end{figure}

To verify that formula~\eqref{eq:app_density_r} provides an accurate representation of the state in configuration space, one can also test the formula on the Coulomb system used in Section \ref{ssec:analytic_test}. The analytical expressions of the radial probability densities for the energy levels $1S$, $2S$ and $1P$ are given by \cite{yane94}
\begin{subequations} \label{eq:coul_analytic_r}
\begin{align}
    &R_{1S}(r)\diff r = \left(2 e^{-r}\right)^2r^2\diff r, \\
    &R_{2S}(r)\diff r = \left(\frac{1}{\sqrt{2}}(1-\frac{r}{2})e^{-r/2}\right)^2r^2 \diff r, \\
    &R_{1P}(r)\diff r = \left(\frac{1}{\sqrt{24}} r e^{-r/2}\right)^2r^2 \diff r.
\end{align}
\end{subequations}
Fig~\ref{fig:comp_densities_r} compares the radial probability densities obtained using equation~\eqref{eq:app_density_r} with the exact ones. One can immediately notice that, for small meshes, the approximate probability densities exhibit unphysical oscillations, especially in the long range sector. This behaviour of the LMM was already reported in References~\cite{lacr11,lacr12}. As the mesh size increases, these oscillations become less pronounced, as shown in Fig~\ref{fig:comp_densities_r}. They can also be mitigated by adjusting the scale parameter $h$. This effect is illustrated in Fig.~\ref{fig:comp_densities_r_bis}, which reproduces Fig.~\ref{fig:comp_densities_r} using a different value of $h$ and displays fewer oscillations. This criterion can also be used to select the value of this parameter. For large meshes, the approximate and exact densities clearly coincide.

\section{Application to constituent model in hadron phenomenology}
\label{sec:app_to_mesons}

In this last section, the LMM in momentum representation is used to study a system of two semi-relativistic particles interacting with Cornell interactions. This model has been used to describe light mesons in Reference~\cite{fulc94}. Consider two particles in their centre-of-mass frame with mass $m_1$ and $m_2$,
\begin{equation}
    T(p^2) = \sqrt{p^2+m_1^2}+\sqrt{p^2+m_2^2},
    \label{eq:rel_kin}
\end{equation}
and that interacts with the following potential,
\begin{equation}
    V(r) = - \frac{\kappa}{r} + a r + C
    \label{eq:fulcher}
\end{equation}
Above, $m_1$, $m_2$, $\kappa$, $a$ and $C$ are phenomenological parameters which are fixed in Reference~\cite{fulc94}. For mesons made of the lightest flavours, they are given by
\begin{align}
    & m_1 = m_2 = 0.150\,\text{GeV}  &&\kappa = 0.437, &&a= 0.203\,\text{GeV}^2, &&C = -0.599\,\text{GeV}.
    \label{eq:param_flu}
\end{align}
Table \ref{tab:fulcher_comp} presents the energies of the $1S$, $2S$ and $1P$ eigenstates obtained by using the LLM in momentum space developed in the current work. Different numbers of mesh points are used. The LMM eigenenergies are compared with those given in the original publication \cite{fulc94} and those from Reference~\cite{sema01}, where a LMM in configuration space has been used to study the same system.

\begin{table}
\centering
\begin{tabular}{l rr r rr r rr}
\hline\hline
$N$ & \multicolumn{2}{c}{$1S$} && \multicolumn{2}{c}{$2S$} && \multicolumn{2}{c}{$1P$}\\ 
\hline
& \hspace{8mm} LMM \cite{sema01}& \hspace{1mm} present LMM &\,\hspace{8mm}\,&  LMM \cite{sema01}& \hspace{1mm} present LMM &\,\hspace{8mm}\,& LMM \cite{sema01}& \hspace{1mm} present LMM \\
\hline
$10$ & $0.702\,373$ & $0.690\,205$ && $1.415\,418$ & $1.499\,120$ && $1.240\,239$ & $1.217\,182$ \\
$20$ & $0.702\,570$ & $0.703\,199$ && $1.415\,854$ & $1.422\,610$ && $1.240\,238$ & $1.237\,518$ \\
$30$ & $0.702\,584$ & $0.702\,660$ && $1.415\,877$ & $1.414\,775$ && $1.240\,238$ & $1.240\,446$ \\
$40$ & $0.702\,587$ & $0.702\,642$ && $1.415\,882$ & $1.416\,084$ && $1.240\,238$ & $1.240\,220$ \\
$50$ & $0.702\,588$ & $0.702\,623$ && $1.415\,884$ & $1.415\,932$ && $1.240\,238$ & $1.240\,240$ \\
$60$ & $0.702\,588$ & $0.702\,614$ && $1.415\,885$ & $1.415\,927$ && $1.240\,238$ & $1.240\,238$\\
$70$ & $0.702\,589$ & $0.702\,609$ && $1.415\,885$ & $1.415\,917$ && $1.240\,238$ & $1.240\,238$ \\
$80$ & $0.702\,589$ & $0.702\,605$ && $1.415\,885$ & $1.415\,911$ && $1.240\,238$ & $1.240\,238$ \\ 
\hline
 \multicolumn{1}{l}{Ref. \cite{fulc94}}&
 \multicolumn{2}{c}{$0.703$}&&
 \multicolumn{2}{c}{$1.416$}&&
 \multicolumn{2}{c}{$1.240$}\\ \hline\hline
\end{tabular}
\caption{Test of the LMM in momentum representation (LMM mom.) for a mesonic system for different mesh sizes $N$. Kinetic and potential energies in the system are given in \eqref{eq:rel_kin} and \eqref{eq:fulcher}. Parameters are given in \eqref{eq:param_flu}. Energies are given in GeV. The eigenenergies obtained in Reference~\cite{sema01} with a LMM in position representation are given in the associate columns. The energy from the original Reference~\cite{fulc94} is given in the last line. Calculations with the LMM in momentum representation are performed with $h=0.5$. \justifying}   
\label{tab:fulcher_comp}
\end{table}

The LMM in momentum space gives good results up to four digits. However, a slower convergence than that of the LMM in configuration space of \cite{sema01} is observed. This is probably because, for simplicity, the calculations in momentum representation depicted in Table \ref{tab:fulcher_comp} use a fixed scale parameter at $h=0.5$, while Reference \cite{sema01} employs a more sophisticated way to choose it (for instance, the latter becomes dependent on the mesh size). For each state, the energy obtained is compatible with that provided in the original Reference \cite{fulc94}. This test illustrates that the LMM in position or momentum representation yields results in close agreement. The user has the freedom to choose the most convenient representation depending on the specificity of its application.

\section{Conclusion}

This paper introduces an alternative way to compute potential matrix elements within a LMM written in momentum space. It mimics the strategy used to handle semi-relativistic kinematics in configuration space \cite{sema01}. This method allows for the resolution of two-body Schrödinger-like equations with any kind of kinematics and potentials, including those encountered in hadron phenomenology. This was not the case with the LMM in momentum space developed in Reference~\cite{lacr12}, which is not appropriate to handle Cornell potentials, for instance. 

The efficiency of the method has been validated by comparison with analytical results for the Coulomb potential and is also compared to that from Reference \cite{lacr12} in \ref{app:gauss}. To illustrate the interest in hadronic physics, the method has been tested on a phenomenological meson model. The corresponding results were compared to those from the original publication~\cite{fulc94} and from Lagrange-mesh calculations in position representation~\cite{sema01}. In addition, special attention has been devoted to manipulating and representing the position and momentum densities, a topic which proved relevant in works such as \cite{buis07,meld08,tour25}.

\section*{Acknowledgements} 

C.C. and J.V. would like to thank the Fonds de la Recherche Scientifique - FNRS for the financial support. The authors thank C. Semay and E. Canivez for their advice and careful reading of the manuscript.


\appendix

\renewcommand{\thesection}{Appendix \Alph{section}}

\renewcommand{\theequation}{\Alph{section}.\arabic{equation}}

\section{Direct calculation of the LMM matrix elements in momentum representation}
\label{app:LMM_direct_calc}

This appendix intends to illustrate the claim that the LMM developed in Reference~\cite{lacr12} cannot be straightforwardly applied to potentials such as the Coulomb or linear one, a significant omission since, as previously noted, these potentials are widely used in hadronic physics to describe the interactions between quarks and gluons in constituent models of QCD. Developments are driven using the Coulomb potential, but similar results can be obtained for a linear one.

The methodology from reference~\cite{lacr12} suggests to evaluate via a direct calculation the potential energy ME,
\begin{equation}
    V_{ij} = \bra{f_i;lm}V(r)\ket{f_j;lm}.
    \label{eq:pot_ME}
\end{equation}
To this end, it aims at evaluating the potential Fourier transform $V_{\text{FT}}(|\bm{p}-\bm{p'}|)$ and decomposes it in partial waves denoted $V_l(p,p')$,
\begin{equation}
    V_{\text{FT}}(k=|\bm{p}-\bm{p'}|) = \frac{1}{2\pi^2k}\int_0^\infty V(r) \sin(kr) r\diff r \quad \text{and} \quad V_l(p,p')= 2\pi\int_{-1}^{+1}P_l(t)V_{\text{FT}}\left(\sqrt{p^2+p'^{\,2}-2pp't}\right)\diff t
\end{equation}
where $P_l$ is a Legendre polynomial \cite{abra64}. Matrix elements~\eqref{eq:pot_ME} are then easily computed using $N$-points Gauss-Laguerre quadratures,
\begin{equation}
    V_{ij} \approx h^3\sqrt{\lambda_i\lambda_j}x_ix_jV_l(hx_i,hx_j).
    \label{eq:pot_ME_formula}
\end{equation}
Altough these formulas prove very convenient to use for short-range potentials, such as the Gaussian or Yukawa ones, it fails when long-range interactions, such as the Coulomb one, are used. In the Coulomb case, $V(r)=-1/r$, one shows that \cite[formula (37) with $b=0$]{lacr12}\cite{hers93}
\begin{equation}
    V_l(p,p') = -\frac{1}{\pi p p'}Q_l\left(\frac{\bar p^2+p^2}{2\bar p p}\right).
\end{equation}
where $Q_l$ is a second kind Legendre function \cite{abra64}. As a result, formula \eqref{eq:pot_ME_formula} becomes
\begin{equation}
    V_{ij} \approx -\frac{h}{\pi}\sqrt{\lambda_i\lambda_j}\,Q_l\left(\frac{x_i^2+ x_j^2}{2 x_i x_j}\right).
    \label{eq:pot_ME_coulomb}
\end{equation}
However, considering the case $i=j$, the argument of the second kind Legendre function becomes identically equal to~$1$, a value for which $Q_l$ diverges. This makes impossible to use formula \eqref{eq:pot_ME_formula} in presence of a Coulomb potential.

The situation for a linear potential is even trickier as, in that case, the potential's Fourier transform necessarily takes a distributional sense. The problem is solved in References~\cite{hers93,norb91}, which allows to write the matrix element~\eqref{eq:pot_ME} in an integral form. However, this integral employs the derivative of a second kind Legendre function, and, as before, the matrix element proves divergent for $i=j$. 

\section{Gaussian potential}
\label{app:gauss}

This appendix aims at comparing results obtained using the present LMM to those obtained with the LMM from Reference~\cite{lacr12}. This comparison is performed on a system of two semi-relativistic particles subjected to a Gaussian potential,
\begin{align}
    &T(p^2) = \sqrt{p^2+m_1^2}+\sqrt{p^2+m_2^2}, &&V(r^2) = -a \exp{(-b^2r^2)},
    \label{eq:kin_pot_comp}
\end{align}
where $m_1$, $m_2$ are the particle's masses and $a$, $b$ are some fixed parameters. Consistently with \cite{lacr12}, these parameters are fixed to 
\begin{align}
    &m_1 = m_2 = 1, &&a=3, &&b=1.\label{eq:param_comp}
\end{align}
With these values, the system accepts a single bound state with energy $E$ such that $0<E<2m_1$.

The Table~\ref{tab:gauss_comp} summarises the comparison. The single bound state energy, as well as various observables, are evaluated. Concerning the results from the present method, the eigenvector is obtained using the methodology depicted in the main text, but observables are evaluated as depicted in Reference~\cite{lacr12}. Although the commentary below equation \eqref{eq:VR2} may suggest that the LMM developed in this paper is more approximate, the test does not conclude in major differences in terms of precision and convergence between the methods\footnote{Let us mention that this test revealed a misprint in Reference \cite{lacr12}: the line name $\braket{\sqrt{p^4}}$ in Table II must be replaced by
$\braket{p^4}$. The misprint has been confirmed by the authors.}. Let us recall that the present LMM can be used to solve more potentials than the LMM from Reference~\cite{lacr12}.

\begin{table}
\centering
\begin{tabular}{lrrrrrr}
\hline\hline
 & \multicolumn{2}{c}{$N=10$} & \multicolumn{2}{c}{$N=20$} & \multicolumn{2}{c}{$N=50$} \\
\hline
 & \hspace{8mm} LMM \cite{lacr12}& \hspace{2mm} Present LMM & \hspace{10mm} LMM \cite{lacr12}& \hspace{2mm} Present LMM&  \hspace{10mm} LMM \cite{lacr12}& \hspace{2mm} Present LMM\\ 
\hline
$E$ & $1.870\,441\,99$ & $1.870\,823\,54$ & $1.871\,008\,78$ & $1.870\,987\,31$ & $1.870\,983\,67$ & $1.870\,983\,62$ \\
$\braket{\sqrt{p^2+m^2}}$ & $1.354\,272\,4$ & $1.353\,714\,5$ & $1.355\,465\,0$ & $1.355\,464\,5$ & $1.355\,380\,7$ & $1.355\,380\,5$\\

$\braket{p^4}$& $3.981\,098$ & $3.975\,794$& $3.992\,369$ & $3.992\,363$& $3.991\,570$ & $3.991\,568$\\ 

$\braket{r}$ & $1.711\,71$ & $1.738\,73$& $1.735\,51$ & $1.732\,95$& $1.733\,76$ & $1.733\,74$ \\

$\braket{U(r)}$ & $-0.838\,109\,4$ & $-0.836\,605\,4$ & $-0.839\,921\,2$ & $-0.839\,941\,6$ & $-0.839\,777\,7$ & $-0.839\,777\,4$ \\
\hline\hline

\end{tabular}
\caption{Comparison of the results obtained with the present LMM in momentum space and that presented in Reference~\cite{lacr12}. The energy eigenvalue and some observables are computed in arbitrary units with both methods for three different mesh sizes. Kinetic and potential energies~\eqref{eq:kin_pot_comp} are used with parameters~\eqref{eq:param_comp}. Same value of $h=0.4$ is chosen for both methods.\justifying}
\label{tab:gauss_comp}
\end{table}


\bibliographystyle{elsarticle-num-names.bst}
\bibliography{ref}


\end{document}